\documentclass[mathleft
]{an}
\usepackage{graphicx}
\usepackage{times}
\begin{document}

\Pagespan{1}{4}
\Yearpublication{2008}%
\Yearsubmission{2008}%
\Month{07}%
\Volume{}%
\Issue{}%

\title{What do dynamical cluster masses really tell us about dynamics?}

\author{Richard de Grijs\inst{1,2}\fnmsep\thanks{\email{R.deGrijs@sheffield.ac.uk}} \and M. B. N. Kouwenhoven\inst{1} \and Simon P. Goodwin\inst{1}}
\titlerunning{Dynamical cluster masses}
\authorrunning{R. de Grijs et al.}
\institute{
Department of Physics \& Astronomy, The University of Sheffield, Hicks Building, Hounsfield Road, Sheffield S3 7RH, UK
\and 
National Astronomical Observatories, Chinese Academy of Sciences, 20A Datun Road, Chaoyang District, Beijing 100012, China}

\received{---}
\accepted{---}
\publonline{---}

\abstract{The diagnostic age versus mass--to--light ratio diagram is
often used in attempts to constrain the shape of the stellar initial
mass function, and the stability and the potential longevity of
extragalactic young to intermediate-age massive star clusters. Here,
we explore the pitfalls associated with this approach and its
potential for use with Galactic open clusters. We conclude that for an
open cluster to survive for any significant fraction of a Hubble time
(in the absence of substantial external perturbations), it is a
necessary but not a sufficient condition to be located close to the
predicted photometric evolutionary sequences for `normal' simple
stellar populations.}


\maketitle

\section{A diagnostic diagram?}

\begin{figure}
\includegraphics[width=\columnwidth]{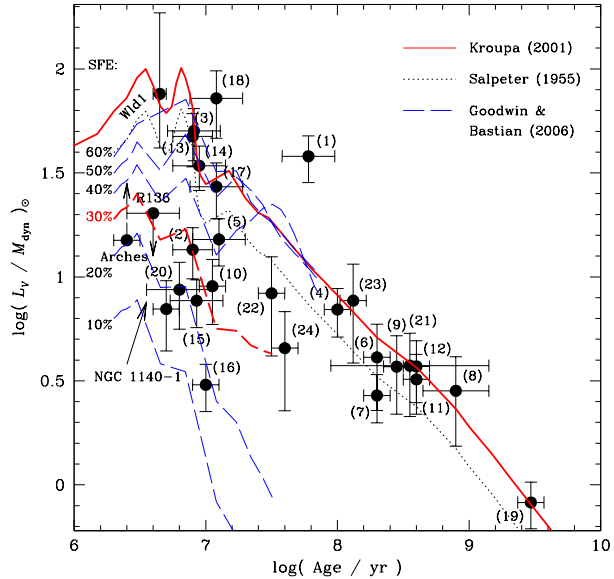}
\caption{Updated version of the $M/L$ ratio versus age diagnostic
   diagram for young and intermediate-age massive star clusters. The
   numbered data points follow de Grijs \& Parmentier (2007; their
   fig. 2); overplotted are the Maraston (2005) SSP predictions for a
   Salpeter (1955) and a Kroupa (2001) stellar IMF, as well as the
   effects of varying effective star-formation efficiencies, based on
   Goodwin \& Bastian (2006).}
\label{mlratio.fig}
\end{figure}

Over the past few years, detailed studies of the stellar content and
longevity of extragalactic massive star clusters have increasingly
resorted to the use of the age versus mass-to-light ($M/L$) ratio
diagram as a diagnostic tool (see Fig. \ref{mlratio.fig} for an
up-to-date version), where one usually compares dynamically determined
$M/L$ ratios with those predicted by the evolution of `simple' stellar
populations (SSPs; e.g., Smith \& Gallagher 2001; Mengel et al. 2002;
McCrady et al. 2003, 2005; Larsen et al. 2004; Bastian et al. 2006;
Goodwin \& Bastian 2006; Moll et al. 2008; see de Grijs \& Parmentier
2007 for a review). Based on high-resolution spectroscopy to obtain
the objects' integrated velocity dispersions, $\sigma$, and on high
spatial resolution imaging to obtain accurate projected half-light
radii, $r_{\rm hl}$, most authors then apply Spitzer's (1987)
equation,
\begin{equation}
\label{spitzer.eq}
M_{\rm dyn} = \eta \frac{r_{\rm hl} \sigma^2}{G} ,
\end{equation}
to obtain the dynamical cluster masses, $M_{\rm dyn}$ ($G$ is the
gravitational constant and $\eta \approx 9.75$ is a dimensionless
parameter which is usually assumed to be constant; but see Fleck et
al. 2006; Kouwenhoven \& de Grijs 2008).

Despite a number of simplifying assumptions (see, e.g., de Grijs \&
Parmentier 2007 for a review; Moll et al. 2008), one can get at least
an initial assessment as to whether a given cluster may be (i)
significantly out of virial equilibrium, in particular `super-virial',
(ii) significantly over- or underabundant in low-mass stars, or (iii)
populated by a significant fraction of binary and higher-order
multiple systems. This has led to suggestions that, in the absence of
significant external perturbations, young massive clusters (YMCs)
located in the vicinity of the SSP models and aged $\ga 10^8$ yr may
survive for a Hubble time and eventually become old globular cluster
(GC)-like objects (e.g., Larsen et al. 2004; Bastian et al. 2006; de
Grijs \& Parmentier 2007).

In this contribution we will first highlight the uncertainties
associated with a few of the YMCs (Section 2) -- and thus question the
use of the velocity dispersion as a tracer of their gravitational
potential -- and then proceed to apply the same diagnostic approach to
a small sample of Galactic open clusters (Section 3). In Section 4 we
will provide a summary, emphasising the pitfalls associated with the
use of dynamical mass estimates as a proxy for cluster longevity.

\section{Closer inspection}

Let us now examine a few of the YMCs from Fig. \ref{mlratio.fig} in
more detail. Using the dynamical mass estimate from Mengel \&
Tacconi-Garman (2007), combined with the integrated photometry of
Piatti et al. (1998), in de Grijs \& Parmentier (2007) we concluded
that the Galactic YMC Westerlund 1 appears to be consistent with a
normal Kroupa (2001) or Salpeter (1955)-type initial mass function
(IMF), despite the significant uncertainties in the observables. Since
in the $V$ band, on which the Piatti et al. (1998) photometry was
based, the confusion between the cluster members and the Galactic
field stellar population is substantial (because of the significant
extinction along this sightline), we obtained imaging observations in
the $I$ band, where this confusion is significantly reduced (de Grijs
et al. 2008).

The combined integrated magnitude of the three brightest red
supergiants, yellow hypergiants, and blue super\-gi\-ants is some 40\%
of the cluster's total integrated $I$-band flux (see de Grijs et
al. 2008 for details). These nine sources stand out from the overall
stellar luminosity function, which appears to otherwise have been
drawn from a `normal' IMF. Each of these sources is in a rare,
short-lived phase and so the luminosity of the cluster might be
expected to vary significantly on short time-scales. Thus, this serves
as a clear caution that stochasticity in the cluster's IMF (e.g.,
Brocato et al. 2000; Barker et al. 2008), as well as stochasticity in
the numbers of stars in unusually luminous post-main-sequence
evolutionary stages (e.g., Cervi\~no \& Valls-Gabaud 2003; Cervin\~no
\& Luridiana 2006) may contribute significantly to variations in a
cluster's $M/L$ ratio.

Thanks to Bastian \& Goodwin (2006) and Goodwin \& Bastian (2006), we
are now confident that we understand why some of the YMCs deviate
significantly from the expected evolution -- for a `standard' IMF --
of the stellar $M/L$ ratio at ages of up to a few $\times 10^7$
yr. However, the location of the massive cluster F in M82 remains to
be understood, despite having been the subject of a considerable
number of recent studies (see, e.g., Bastian et al. 2007 for an
overview).

Smith \& Gallagher (2001) concluded that, assuming that their age ($60
\pm 20$ Myr), mass ($1.2 \pm 0.1 \times 10^6 M_\odot$) and luminosity
($M_V = -14.5 \pm 0.3$ mag; $L_V / M_{\rm dyn} = 45 \pm 13 \;
L_{V,\odot} / M_\odot$) measurements for the object were correct, it
must either have a low-mass cut-off of its mass function at $\sim 2-3
M_\odot$, or a shallow mass function slope, $\alpha \sim 2$ for a mass
function including stellar masses down to $0.1 M_\odot$. McCrady et
al. (2003) suggested, based on near-infrared population synthesis
modeling, that Smith \& Gallagher (2001) had overestimated their ages
as well as the half-mass radius of the cluster; they derived the
latter more accurately based on higher-resolution imaging. However, in
a recent paper Bastian et al. (2007) used optical spectroscopy,
confirming the YMC's age to be in the range of 50--70 Myr. However,
contrary to the earlier studies, Bastian et al. (2007) conclude that
M82-F is subject to a large amount of differential extinction, thus
rendering earlier luminosity estimates based on the assumption of
foreground-screen extinction very uncertain. They conclusively show
that the apparently large degree of mass segregation in the cluster,
derived by McCrady et al. (2005) based on the variation of $r_{\rm hl}$
as a function of wavelength, is in fact caused by this differential
extinction. Applying their spatially resolved extinction corrections
to {\sl Hubble Space Telescope} imaging data, Bastian et al. (2007)
also suggest that the original cluster size reported by Smith \&
Gallagher (2001) was underestimated. Using their updated (larger) size
estimate, the discrepant M82-F data point moves closer to the SSP
predictions, hence suggesting that the cluster may be less enigmatic
than previously thought.

Finally, we single out YMC NGC 1140-1 (Hunter et al. 1994; de Grijs et
al. 2004; Moll et al. 2008) for a closer look. There are various
potential explanations for the difference between its observed and the
SSP model $M/L_V$ ratio. Assuming that this difference is due solely
to the cluster being out of virial equilibrium, its effective
star-formation efficiency (eSFE; Goodwin \& Bastian 2006) is around
10--20\%. The eSFE is a measure of the extent to which a cluster is
out of equilibrium after gas expulsion, on the basis that the virial
ratio {\it immediately before} gas expulsion was $Q_{\rm vir} =
T/|\Omega| = 1/2$(eSFE), where $T$ and $\Omega$ are the kinetic and
potential energy of the stars, respectively, and a system in virial
equilibrium has $Q_{\rm vir} = 1/2$. The eSFE corresponds to the true
SFE if the stars and gas were initially in virial equilibrium. Goodwin
\& Bastian (2006) predict that a cluster with an eSFE of 10\% will
lose 80\% of its mass in 20 Myr and not remain bound. As such, the
cluster would disperse over a relatively short time-scale and,
therefore, not evolve into a second-generation GC. More likely,
however, the crowded nature of the region may have caused the cluster
velocity dispersion (and hence the $M/L$ ratio) to be
overestimated. The broadening of the red-supergiant features seen in
the knot spectra may not have been due solely to the virial motions of
the brightest cluster within the knot, as was assumed, but may have
had contributions from the motions of nearby contaminating clusters
instead (cf. Moll et al. 2008).

\begin{figure}
\includegraphics[width=\columnwidth]{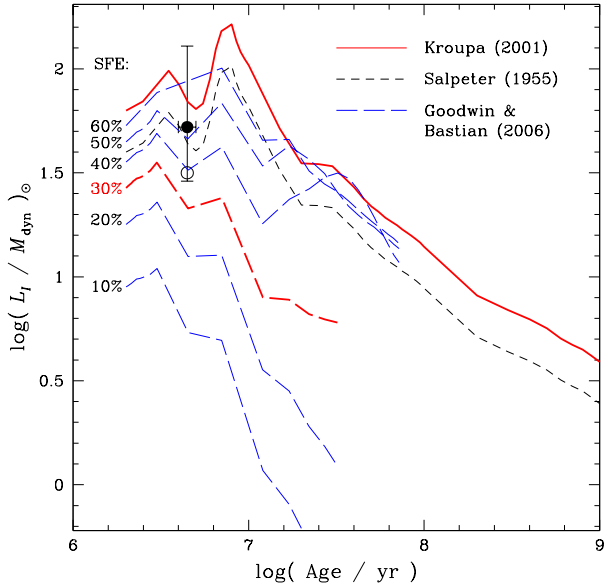}
\caption{Westerlund 1 in the diagnostic age versus $M/L_V$ ratio
diagram. The open circle represents the shift in the cluster's locus
from its current position (solid bullet) if we were to exclude the
nine brightest stars; this exemplifies the uncertainties introduced by
stochastic IMF sampling and by having fortuitously caught the cluster
at a time when it is dominated by a small number of very bright
stars.}
\label{wld1}
\end{figure}

\section{Galactic open clusters}

\begin{figure}
\includegraphics[width=\columnwidth]{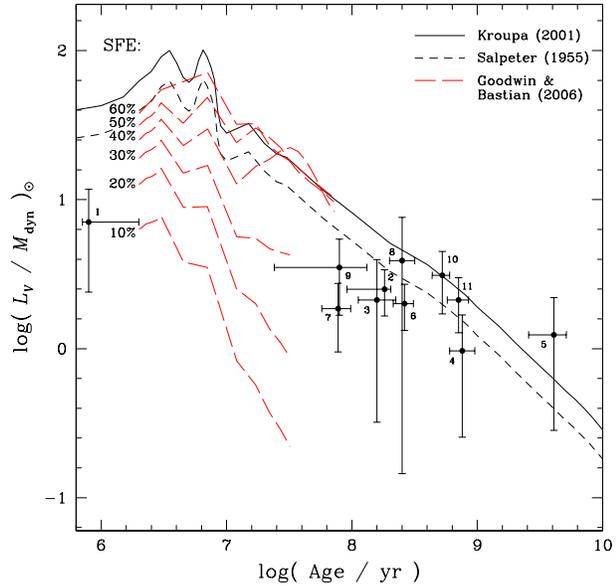}
\caption{As Fig. 1, but for those Galactic open clusters for which
velocity dispersion measurements are available. Numbered clusters: 1,
NGC 1976 (Orion Nebula Cluster); 2, NGC 2168; 3, NGC 2516; 4, NGC
2632; 5, NGC 2682; 6, NGC 3532; 7, NGC 5662; 8, NGC 6705; 9, Pleiades;
10, Coma Berenices; 11, Hyades.}
\label{ocl.fig}
\end{figure}

We will now explore whether we can also use the same diagnostic
diagram to assess the stability, formation conditions, binarity and
longevity of those open clusters in the Milky Way for which the
requisite observational data exist in the literature. Using the
observational data compiled in de Grijs et al. (2008), we applied
Spitzer's (1987) equation, Eq. (\ref{spitzer.eq}), to derive the
dynamical masses for each of our sample clusters, and then calculated
their $M/L_V$ ratios. The clusters' loci in the diagnostic diagram are
shown in Fig. \ref{ocl.fig}.

We have also included the expected evolution of clusters formed with a
variety of eSFEs. Owing to the nature of our sample,
only\footnote{Despite the extent of the error bar associated with the
age estimates of the Pleiades, Kroupa et al. (2001) showed this
cluster to have re-virialised by an age of 50 Myr, so that it is
unlikely affected by the aftermath of the gas-expulsion phase.} the
Orion Nebula Cluster (ONC; cluster 1 in Fig. \ref{ocl.fig}) is young
enough so as to possibly be affected by the effects of rapid gas
expulsion, as shown by the extent (in terms of age) of the long-dashed
lines in Fig. \ref{ocl.fig}. The majority of our sample clusters are
old enough ($\ga 40$ Myr) to have re-virialised after gas
expulsion. The dynamical state of these objects is therefore dominated
by the combined effects of (internal) two-body relaxation, binary
motions, and external perturbations.

We note that, despite the sometimes significant uncertainties
(parameter ranges) associated with the individual quantities required
to calculate the dynamical $M/L_V$ ratios, the sample clusters follow
the general trend predicted by the SSP models rather closely. The fact
that these clusters lie close to (although systematically somewhat
below) the SSP predictions should not be a suprise. Clusters
significantly below the SSP lines are dynamically `hot' and expected
to dissolve rapidly, whilst clusters significantly above the lines
will be dynamically `cold' and should (re-)virialise over a few
crossing times to move closer to the canonical SSP lines.

Most of the sample clusters are found somewhat below the SSP model
curves. Errors in the core radii are expected to be random, and
unbiased by the mass of a cluster. However, the use of the {\it core}
velocity dispersions and radii may introduce a systematic bias in the
dynamical mass estimates. The majority of the star clusters in the
sample are older than $\sim 10^8$ yr, which implies that they have
ages greater than their (initial) half-mass relaxation times
(cf. Danilov \& Seleznev 1994). Therefore, these clusters are expected
to be close to energy equipartition, and thus are significantly mass
segregated. Equipartition reduces the {\it global} velocity dispersion
of high-mass stars relative to low-mass stars, causing high-mass stars
to migrate to the cluster core. Therefore, we might expect the core
velocity dispersion of low-mass cores to underestimate the dynamical
mass and thus produce colder clusters -- as observed.

Alternatively, the effect of binaries within clusters may well account
for most of the displacement of the observed cluster positions below
the model curves. Kouwenhoven \& de Grijs (2008) pointed out that if
the velocity dispersion of binary systems was similar to the velocity
dispersion of the cluster as a whole, the {\it observationally
measured} velocity dispersion would overestimate the mass of a
cluster. Based on a comparison with Kouwenhoven \& de Grijs (2008), it
appears that the vast majority of the sample clusters are indeed
expected to be binary dominated (see de Grijs et al. 2008 for
details).

We also note that the cluster masses may well have been overestimated
by factors of a few through the universal use of
Eq. (\ref{spitzer.eq}). In particular, for highly mass-segregated
clusters containing significant binary fractions, a range of stellar
IMF representations, and for combinations of characteristic relaxation
time-scales and cluster half-mass radii, the adoption of a single
scaling factor $\eta \approx 9.75$ introduces systematic offsets. To
correct for these, we would need to adopt smaller values of $\eta$
(e.g., Fleck et al. 2006; Kouwenhoven \& de Grijs 2008), and this
would thus lead to dynamical mass overestimates if $\eta = 9.75$ were
assumed.

In addition, the mass functions (MFs) of clusters will be altered due
to the preferential loss of low-mass stars during two-body
relaxation. This will result in an increasingly `top-heavy' MF in
clusters, which will raise them to lower $M/L_V$ ratios than would be
expected from the canonical SSP models. The degree to which the MF
will change depends on the two-body relaxation time which, to first
order, depends on the mass of the cluster (and also on its size;
however, we ignore this for now). Thus, low-mass clusters are expected
to have top-heavy MFs as compared to high-mass clusters. Therefore,
one would expect low-mass clusters to lie some way above the canonical
SSP models, and high-mass clusters to lie slightly above these
lines. However, the effect of binaries in clusters is to increase the
observed velocity dispersion and so overestimate the masses of
clusters, shifting them back down towards, and even below, the
canonical SSP lines.

\section{Concluding remarks}

In this contribution we have explored the usefulness of the diagnostic
age versus $M/L$-ratio diagram in the context of YMCs and Galactic
open clusters. This diagram is often used in the field of
extragalactic young to intermediate-age massive star clusters to
constrain the shape of their stellar IMF, as well as their stability
and the likelihood of their longevity. Using the massive young
Galactic cluster Westerlund~1 as a key example, we caution that
stochasticity in the IMF introduces significant additional
uncertainties. Therefore, the stability and long-term survival chances
of Westerlund~1 remain inconclusive. Similarly, we have highlighted
observational issues associated with M82-F and NGC 1140-1 which render
the use of their velocity dispersions as a diagnostic tool
questionable at best.

For low-mass open clusters it seems clear that the effect of binaries,
mass segregation, and the dynamical alteration of mass functions by
two-body relaxation are important factors that cannot be ignored. Most
importantly, however, we conclude that for an open cluster to survive
for any significant length of time (in the absence of substantial
external perturbations), it is a necessary but not a sufficient
condition to be located close to the predicted photometric
evolutionary sequences for `normal' SSPs. This is based on two of the
sample clusters, the Coma Berenices cluster (number 10 in Fig. 3) and
the Hyades (11), which are known to be in a late stage of dissolution,
yet lie very close indeed to either of the evolutionary sequences
defined by the Salpeter (1955) or Kroupa (2001) IMFs.

\acknowledgements We are grateful for helpful criticism from and
useful discussions with Pavel Kroupa and Mark Gieles. MK gratefully
acknowledges the organising committee for financial support to attend
the meeting. We also acknowledge research support and hospitality at
the International Space Science Institute in Bern (Switzerland), as
part of an International Team programme.

\end{document}